\documentclass{jpsj-suppl}
\usepackage{txfonts} 
\usepackage{slashed}
\usepackage{subfigure}

\title{Threshold Neutral Pion Photoproduction on the Proton}

\author{Astrid \textsc{Hiller Blin}$^{1}$, Manuel \textsc{Vicente Vacas}$^{1}$ and Tim \textsc{Ledwig}$^{1}$}

\inst{$^{1}$Departamento  de  F\'isica  Te\'orica  and  IFIC,  Centro  Mixto  Universidad  de  Valencia-CSIC,
Institutos  de  Investigaci\'on  de  Paterna,  E-46071  Valencia,  Spain}

\email{astrid.blin@ific.uv.es}

\abst{The neutral pion photoproduction on the proton near threshold has a very small scattering cross section when 
compared to the charged channels, which in ChPT is explained by strong cancellations between the lowest order pieces. 
Therefore it is very sensitive to higher-order corrections of chiral perturbation 
theory. We perform a fully covariant calculation up to chiral order $p^3$ and 
we investigate the effect of the inclusion of the $\Delta(1232)$ resonance as an explicit degree of freedom. We 
show that the convergence improves, leading to a much better agreement with data at a wide range of 
energies.}

\kword{neutral pion photoproduction, covariant ChPT, $\Delta(1232)$ resonance}

\begin{document}
\maketitle

\section{Introduction}

There have been many studies of the pion photoproduction data near threshold in the past few decades
~\cite{Bernard:1991rt,Bernard:1992nc,DeBaenst:1971hp,Vainshtein:1972ih,Drechsel:1992pn,Bernard:2006gx,Bernard:1994gm,Bernard:1995cj,Bernard:2001gz}, which were in good agreement with the data existing at the time~\cite{Mazzucato:1986dz,Beck:1990da}. While 
ChPT models very well describe the charged channels of pion photoproduction even at tree level, this is not 
the case for the neutral channels, where the inclusion of higher orders and loop diagrams leads to important 
contributions. This is due to the strong cancellations between amplitude pieces at lowest order, 
which makes the cross sections sensitive to even small corrections. 

We focus on the study of the neutral pion production channel off the proton. New data from the Mainz Microtron~\cite{Hornidge:2012ca} 
have shown that the existing models fail to converge at $\mathcal{O}(p^3)$ for energies higher than 
approximately 20~MeV above threshold, both in a fully covariant approach and in the non-relativistic heavy-baryon 
approximation~\cite{FernandezRamirez:2012nw,Hilt:2013uf}. The situation is aggravated by the fact that the corrections coming 
from the fourth order are very small, although they mean the inclusion of many new free parameters, the low-energy 
constants of ChPT.

Our strategy is to stay at chiral order $p^3$ and to include the $\Delta(1232)$ resonance as an explicit degree of 
freedom. This choice is due to the fact that, since the total cross section of the $\pi^0$ production off the proton 
is very small close to threshold, the $\Delta(1232)$ resonance could have a quite important contribution~\cite{Hemmert:1996xg}, whose effect becomes 
ever clearer when moving to energies closer to its mass~\cite{Ericson:1988gk}. The effect of the inclusion of this resonance has been thoroughly studied in~\cite{Alarcon:2012kn,Chew:1957tf,Adler:1968tw,Pascalutsa:2006up,Pascalutsa:2004pk,Pascalutsa:2005vq,FernandezRamirez:2005iv,Lensky:2009uv}. We show that, although we are not including any additional fitting 
constants, the dynamics of the $\Delta(1232)$ resonance is a sufficient addition to well describe the experimental input 
even at photon energies higher than 200~MeV.

\section{The ChPT Lagrangian}

We describe the pion production in a fully covariant $\mathcal{O}(p^3)$ ChPT calculation including 
nucleons, pions and photons. In this case, the needed terms of the Lagrangian read
\begin{align}
\nonumber\mathcal{L}_{\pi N}=\bar{\Psi}\Bigg\{
&\mathrm{i}\slashed{\mathrm{D}}-m+\frac{g_A}{2}\slashed{u}\gamma_5
+\frac{1}{8m}\left(
c_6 f_{\mu\nu}^+ + c_7\text{Tr}\left[f_{\mu\nu}^+\right]
\right)\sigma^{\mu\nu}
\\
\nonumber&+\left(\frac{\mathrm{i}}{2m}\varepsilon^{\mu\nu\alpha\beta}\left(d_8\text{Tr}\left[\tilde f_{\mu\nu}^+u_\alpha\right]
 +d_9\text{Tr}\left[f_{\mu\nu}^+
\right]u_\alpha
\right)+\text{h.c.}\right)
\mathrm{D}_\beta
\\
 &+\frac{\gamma^{\mu}\gamma_5}{2}\left(
d_{16}\text{Tr}\left[\chi_+\right]u_\mu
 +\mathrm{i}d_{18}\gamma^{\mu}\gamma_5[\mathrm{D}_\mu,\chi_-]
\right)
\Bigg\}\Psi+\dots,
\label{eqLag12}
\end{align}
with the definitions given in~\cite{Fettes:2000gb}.
Furthermore, we include the degrees of freedom of the isospin-3/2 $\Delta$ quadruplet. The relevant terms for the calculations up to the considered 
order in this paper are
\begin{align}
\mathcal{L}_{\Delta\pi N}=\mathrm{i}\bar{\Psi}\left(\frac{h_A}{2FM_\Delta}T^a\gamma^{\mu\nu\lambda}(\partial_\lambda\pi^a)
+\frac{3\mathrm{i}eg_M}{2m(m+M_\Delta)}T^3\tilde{F}^ {\mu\nu}\right)\partial_\mu\Delta_\nu + \text{H.c.}
\end{align}
The definitions can be found in~\cite{Pascalutsa:2006up} and the full set of diagrams in~\cite{Blin:2015}.

The power-counting scheme we follow here is the so-called $\delta$ expansion, where the mass difference 
$\delta = M_\Delta - m$ is treated as being an additional small quantity of order $p^{1/2}$. This counting is 
adequate for energies close to the pion production threshold and sufficiently far from the $\Delta(1232)$ resonance. The order 
of a diagram is then given by~\cite{Pascalutsa:2002pi}
\begin{equation}
D=4L+\sum{kV_{k}}-2N_\pi-N_N-\frac{1}{2}N_\Delta.
\label{eqOrder}
\end{equation}
When doing this expansion, the loop diagrams with virtual isospin-$3/2$ states start appearing only at order $p^{7/2}$. Therefore, for a $\mathcal{O}(p^3)$ 
calculation it is sufficient to consider the virtual-$\Delta(1232)$ tree diagrams only.

While ChPT has a clear order-by-order counting when treating mesons only, the inclusion of baryons into the theory 
leads to the emergence of power-counting breaking terms. They appear in diagrams where baryon propagators 
are inside of a loop. After integrating over the virtual loop momenta, terms of a lower order than the nominal diagram order 
might appear. In addition, the usual divergences of dimensional regularization also have to be dealt with. We opt for the 
EOMS regularization scheme, where both these problems are addressed simultaneously, by absorbing their fully analytical 
expressions into the low-energy constants of the corresponding order~\cite{Gegelia:1999gf,Fuchs:2003qc}. The success 
of this renormalization scheme has been thoroughly studied on many physical observables in~\cite{Fuchs:2003ir,Lehnhart:2004vi,Schindler:2006it,Schindler:2006ha,Geng:2008mf,Geng:2009ik,MartinCamalich:2010fp,Alarcon:2011zs,Ledwig:2011cx,Alvarez-Ruso:2013fza,Chen:2012nx,Ledwig:2014rfa}. 
More specifically, we use the modified minimal subtraction $\widetilde{MS}$ scheme~\cite{Scherer:2012xha}.

\section{Results and discussion}

\begin{figure}[tbh]
\includegraphics[width=.33\textwidth]{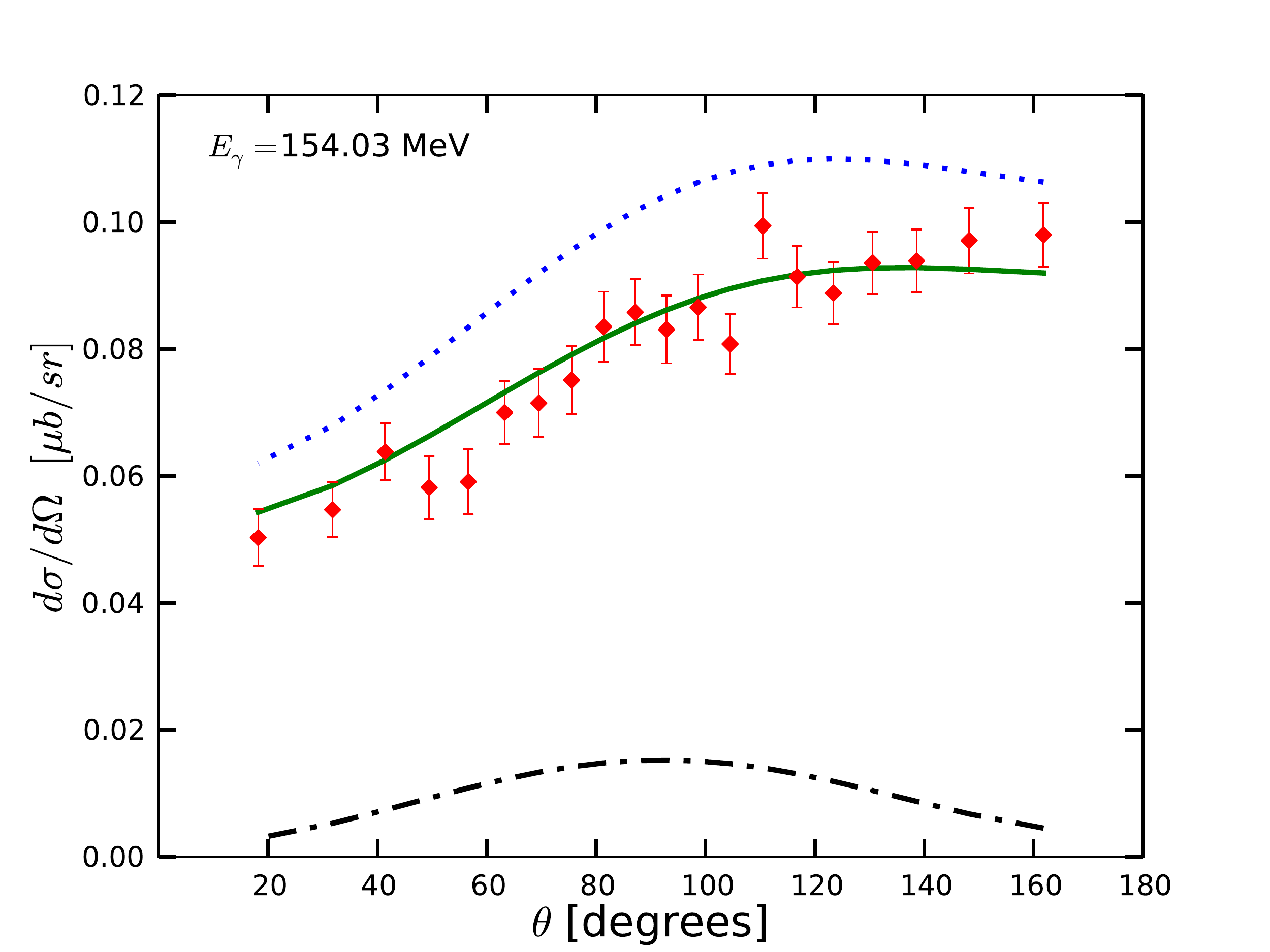}
\includegraphics[width=.33\textwidth]{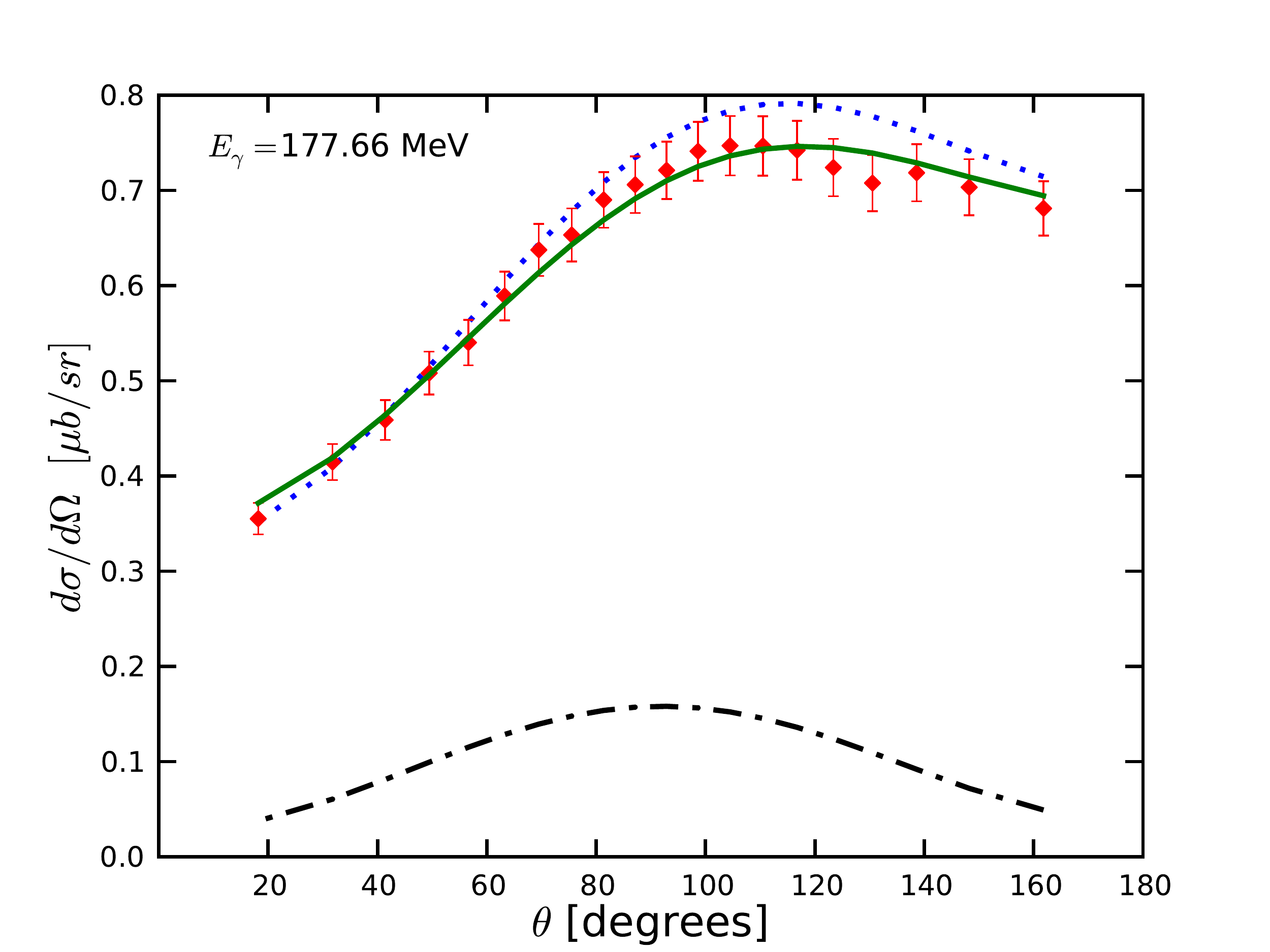}
\includegraphics[width=.33\textwidth]{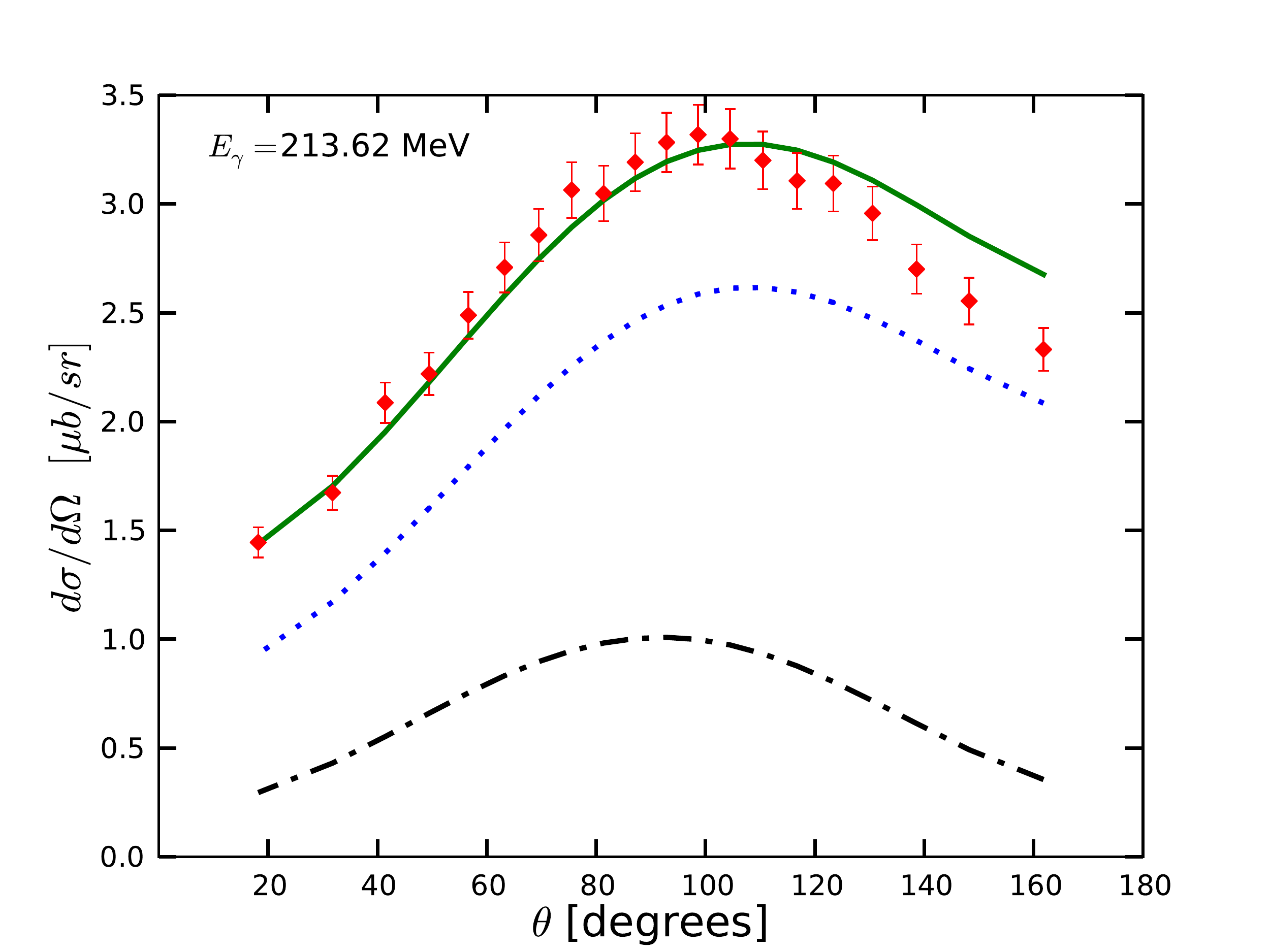}
\caption{Comparison between the theoretical model without (blue dotted line) and with (green line) the inclusion of the $\Delta(1232)$ resonance, at three different energies. The size of the contribution of the diagrams with a virtual isospin-$3/2$ state only is also shown as the black dash-dotted line. The data points are taken from~\cite{Hornidge:2012ca}.}
\label{fDXS}
\end{figure}

\begin{figure}[tbh]
\begin{minipage}{6in}
  \centering
  \raisebox{-0.5\height}
{\subfigure[]{\includegraphics[width=.47\textwidth]{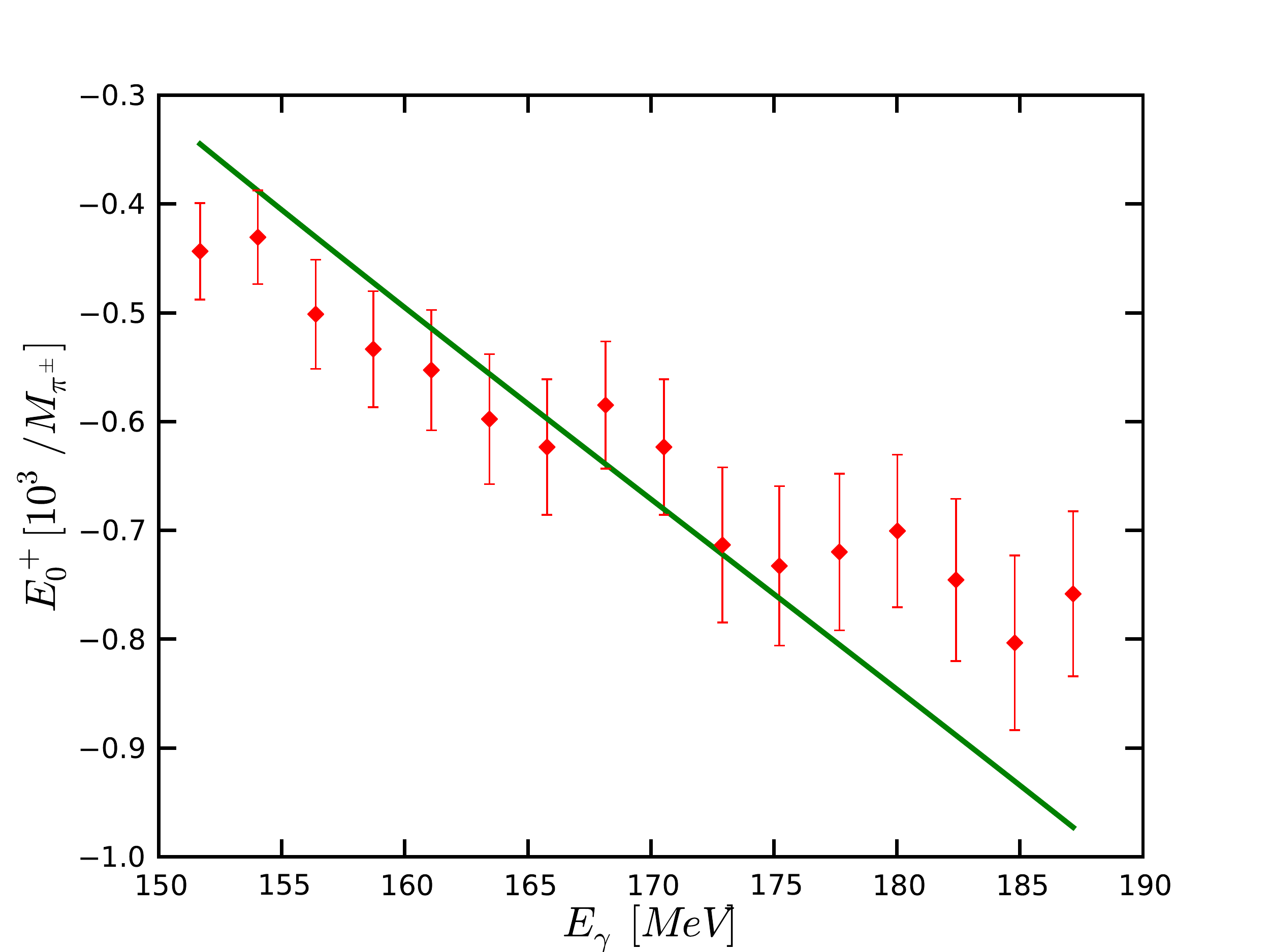}}}
  \hspace*{.2in}
  \raisebox{-0.5\height}{
\subfigure[]{
\label{fO1a}\includegraphics[width=.47\textwidth]{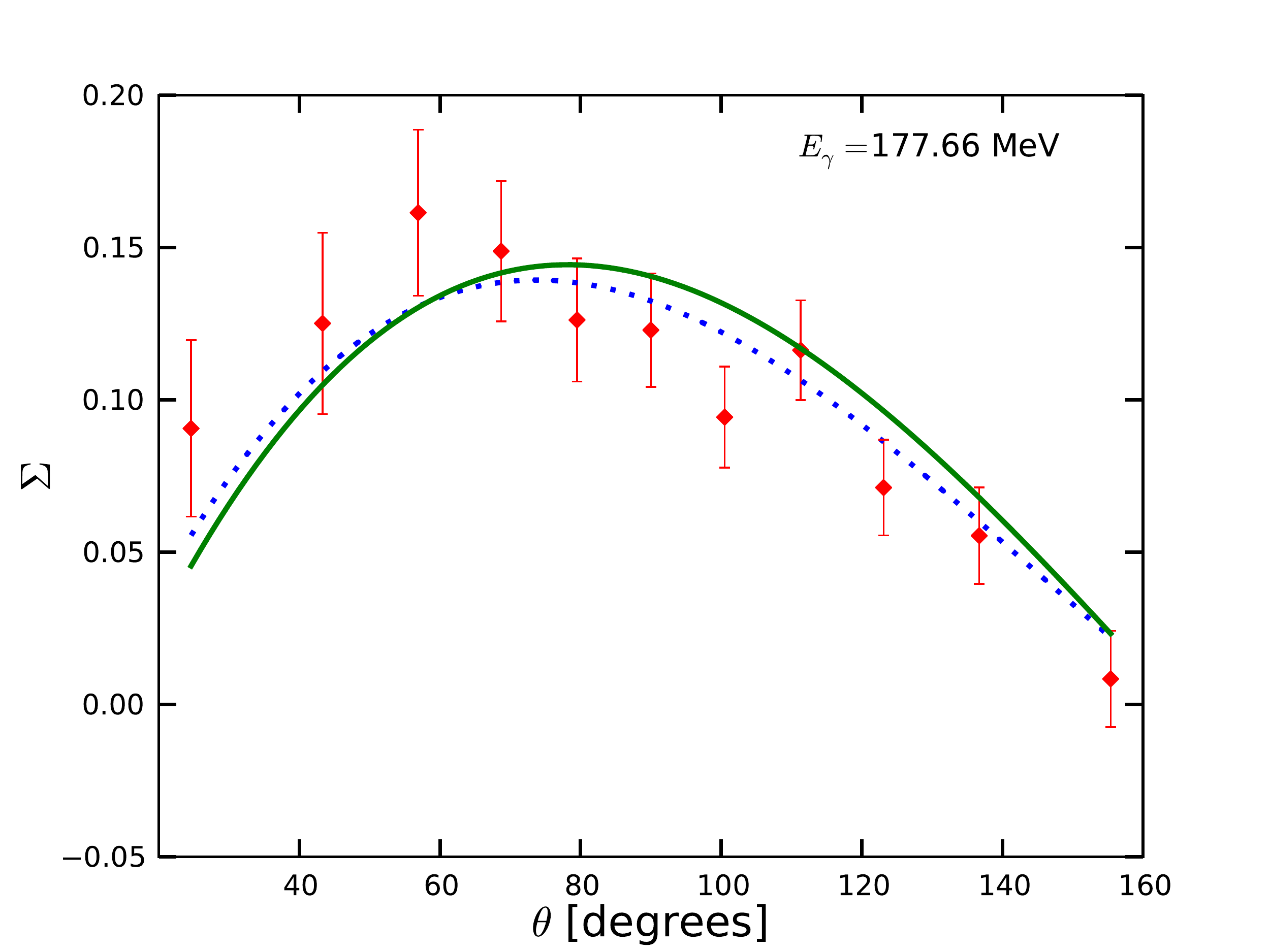}}}
\end{minipage}
\caption{(a) Real part of the multipole $E_0^+$; (b) Photon asymmetries for the model without (blue dotted line) and with (green line) the $\Delta(1232)$ resonance. The data points are taken from~\cite{Hornidge:2012ca}.}
\label{fOtherObserv}
\end{figure}

With the model introduced in the previous section, we did a $\chi^2$ fit to the data from~\cite{Hornidge:2012ca}. At first, we tested the behaviour when not including the spin-$3/2$ states. We could qualitatively reproduce the results already found in~\cite{Hilt:2013uf}: The cross sections are overestimated near threshold and underestimated at high energies. This is because one is missing the component which  describes the steepness of the variation of the cross-section size with the energy. It is solved by including the $\Delta(1232)$-resonance dynamics, although no new fitting constants are introduced. The results are depicted in Fig.~\ref{fDXS}. Although the size of the diagrams with virtual isospin-$3/2$ states is actually very small, they are essential to well describe the data with ChPT. For completeness, we also show the comparison between data and theoretical model of other observables at some example energies in Fig.~\ref{fOtherObserv}.

As future prospects, we want to study the effects of the inclusion of the next-order diagrams, which would introduce loop diagrams with virtual isospin-$3/2$ states. Furthermore, the aim of this study was to show how, when restricting the fitting constants in the exact same way as done previously in models without including the $\Delta(1232)$ resonance, the theoretical model converges much better even at high energies. In order to be able to make statements about the values of the low-energy constants, it is necessary to extend this work to other pion-production channels, so being able to have more data and to study different combinations of the low-energy constants.

\section*{Acknowledgements}
This research was supported by the Spanish Ministerio de Econom\'ia y Competitividad and European FEDER funds under Contracts No. FIS2011-28853-C02-01 and FIS2014-51948-C2-2-P, by Generalitat Valenciana under Contract No. PROMETEO/20090090 and by the EU HadronPhysics3 project, Grant Agreement No. 283286. A.N. Hiller Blin acknowledges support from the Santiago Grisol\'ia program of the  Generalitat Valenciana. We thank D. Hornidge
for providing us with the full set of data from Ref.~~\cite{Hornidge:2012ca}.


\end{document}